\documentclass[12pt,a4paper,english,numbers=noenddot]{article}

\usepackage[top=30truemm,bottom=30truemm,left=25truemm,right=25truemm]{geometry}

\usepackage{amsmath}             % advanced math extension
\usepackage{amssymb}             % new math symbols
\usepackage{amsthm}              % theoremstyle und proof-Umgebung
\usepackage{amsfonts}
\usepackage{array}
\usepackage{color}              % coloured text
\usepackage{graphicx}            % external pictures
\usepackage{lmodern}
\usepackage[automark]{scrpage2}
\usepackage[T1]{fontenc}
\usepackage{algorithm}
\usepackage[noend]{algorithmic}
\usepackage{cite}
\usepackage{rotating}
\usepackage{makeidx}
\usepackage{titlesec}
\usepackage{lscape}
\usepackage{threeparttable}
\usepackage{calc}
\usepackage{multicol}
\usepackage{enumitem}
\usepackage{chemarrow}
\usepackage{dcolumn}
\usepackage{url}

\theoremstyle{definition}

\newtheorem*{theorem*}{定理}

\newtheorem*{definition*}{定義}
\newtheorem*{remark*}{Remark}

\definecolor{commentcolor}{rgb}{0.6, 0.6, 0.6}

\makeatother

\usepackage{babel}
\begin{document}
%\input{nocites}
%\conferenceinfo{ISSAC2012,} {July 22--25, 2012, Grenoble, France.}
%\CopyrightYear{2012} \crdata{XXX} \clubpenalty=10000 \widowpenalty = 10000

\title{An efficient reduction strategy for signature-based algorithms to compute Gr\"obner basis}

\author{%
%{\normalsize 
  Kosuke Sakata\footnote{I would like to thank Professor Shushi Harashita for his helpful comments.}
\footnote{I am grateful to Professor Kazuhiro Yokoyama and Professor Masayuki Noro for discussions on the topic of this manuscript.}\\
  Graduate School of Environment and Information Sciences\\
Yokohama National University\\
Yokohama Kanagawa Japan \\
kosuke-sakata-tc@ynu.ac.jp\\
%}
}
\maketitle

%\category{I.1.2}{Symbolic and Algebraic Manipulation}{Algorithms}[Algebraic Algorithms]
%\category{F.2.1}{Analysis of Algorithms and Problem Complexity}{Numerical Algorithms and Problems}[Computations on Polynomials]
%\terms{Algorithms}

%testing git right now!
%\keywords{Gr\"obner bases, F5 Algorithm, $\ggv$\ Algorithm, incremental signature-based
%  algorithms}

\begin{abstract}
This paper introduces a strategy for signature-based algorithms to compute Gr\"obner basis.
The signature-based algorithms generate S-pairs instead of S-polynomials, and use $\mathfrak{s}$-reduction instead of the usual reduction used in the Buchberger algorithm.
%S-pairs (analogy of S-polynomials) are reduced with the procedure named $\mathfrak{s}$-reduction.
%The $\mathfrak{s}$-reduction has two strategies: 
There are two strategies for $\mathfrak{s}$-reduction: 
one is the only-top reduction strategy which is the way that only leading monomials are $\mathfrak{s}$-reduced. 
The other is the full reduction strategy which is the way that all monomials are $\mathfrak{s}$-reduced.
A new strategy, which we call selective-full strategy, for $\mathfrak{s}$-reduction of S-pairs is introduced in this paper.
%It enables to decrease the number of times of reductions. 
In the experiment, this strategy is efficient for computing the reduced Gr\"obner basis.
For computing a signature Gr\"obner basis, it is the most efficient or not the worst of the three strategies. 

\end{abstract}

\section{Introduction}

The Gr\"obner basis algorithm using signature was first introduced by Faug{\`e}re\cite{F5}.
The algorithm, F5, decreases the number of times of reducing S-pairs (an analogy of S-polynomials) to zero by removing useless critical pairs comparing to existing other algotrithms.
After that, several signature-based algorithms to compute Gr\"obner basis were introduced such that F5C\cite{F5C}, GVW\cite{GVW} and so on. 
The paper\cite{Eder17} by Eder and Faug{\`e}re compiled the thesis related to signature-based algorithms.
We can overview the signature-based algorithms by the paper.

%It is the same for the proposed algorithms that they $\mathfrak{s}$-reduce S-pairs after generating S-pairs.
In the signature-based algorithms, for removing useless critical pairs, S-pairs are not reduced like the Buchberger algorithm, but reduced with more restrictions.
That restricted reduction, called $\mathfrak{s}$-reduction, is indispensable in the algorithms.
There are two strategies for $\mathfrak{s}$-reducing S-pairs according to the paper\cite{Eder17}.
%The paper\cite{Eder17} shows the two strateies for $\mathfrak{s}$-reducing S-pairs. 
One is the only-top reduction strategy ({\bf Algorithm 3}): 
after generating an S-pair, regular $\mathfrak{s}$-reduce leading monomials until the leading monomial cannot be regular $\mathfrak{s}$-reduced. 
The other is the full reduction strategy ({\bf Algorithm 4}): 
after generating an S-pair, regular $\mathfrak{s}$-reduce the all monomials included in the S-pairs.
It cannot be said that either is efficient because it depends on polynomial systems we solve and on strategies we use.

For efficient Gr\"obner basis computation, to decrease the number of reductions is one of the significant problems, because the reduction process accounts for a large propotion in the computation.
This paper introduces a new strategy for $\mathfrak{s}$-reduction aiming to decrease the number of sum of $\mathfrak{s}$-reductions and usual reductions.
Overview of the starategy is following: 
after generating an S-pair, we fulfill only-top reduction. 
If the S-pair meets a certain condition(\fbox{{\bf SF}} in \S 4), we execute full reduction.
We name the strategy the {\it selective-full reduction strategy}({\bf Algorithm 5}).
Efficiency of the strategy was evaluated by some Gr\"obner basis benchmarks.
The selective-full strategy process fewer times of reduction for computing the reduced Gr\"obner basis.
For computing a signature Gr\"obner basis, it is the most efficient or not the worst of the three strategies. 
%It decreases the number of sum of $\mathfrak{s}$-reductions and usual reductions for computing Gr\"obner basis for most bechmarks.
%Especially, for computing reduced Gr\"obner basis, the proposed strategy is the most efficient in the three strategies.

\section{Notation}
In this section, we review the definitions around the signature-based algorithms.
Let $R$ be a polynomial ring over a field $K$, let ${f_1, f_2, \dots, f_m} \in R$, let ${\bf e}_1,{\bf e}_2,\dots,{\bf e}_m$ be the standard basis of a free module $R^m$.
Consider the following homomorphism

$$
\begin{array}{ccccc}
\bar{{}}:&R^m & \longrightarrow & R & \\ \\
&\alpha = {\displaystyle \sum_{i=1}^{m}} a_i {\bf e}_i  & \longmapsto & \overline{\alpha} ={\displaystyle \sum_{i=1}^{m}} a_i f_i , \\
\end{array}
$$

\noindent
where $a_1,\dots,a_m \in R$, especially $\overline{{\bf e}_i} = f_i$.

We choose a monomial order $\leq$ on $R$, and choose a module order $\preceq$ which is compatible with the monomial order, namely we require that for all monomials $a,b \in R$, $a \leq b$ if and only if $a {\bf e}_i \preceq b {\bf e}_i$ for $i=1,\dots,m$.
%Let $\leq$ be a monomial order on $R$.
%We also use the same symbol $\leq$ a module order on $R^m$, as it would not confuse us.
%We choose a monomial order and a module order which is compatible with the monomial order as following: for all monomials $a,b \in R$ and $i=1,\dots,m$, $a \leq b \Leftrightarrow a {\bf e}_i \leq b {\bf e}_i$.
An element of $R^m$ of the form $a {\bf e}_i$ for a monomial $a$ of $R$ is called a {\it term} of $R^m$.
The following POT(position over term) order is one of examples of module orders: let $a {\bf e}_i$, $b {\bf e}_j$ be two module terms in $R^m$, $a {\bf e}_i \preceq_{\operatorname{POT}} b {\bf e}_j$ if and only if either $i<j$ or $i=j$ and $a<b$.
For $\alpha \in R^m$, the {\it signature} $\mathfrak{s}(\alpha)$ of $\alpha$ is defined to be the leading term of $\alpha$ with respect to the module order.
For $f \in R$,we denote by $\operatorname{LT}(f)$ of $f$ the leading term with respect to the monomial order.
We denote by $\operatorname{T}(f)$ the set of the monomials in $f$.

The S-{\it pair} of $\alpha, \beta \in R^m$ is defined to be
\begin{eqnarray}
\operatorname{spair}(\alpha,\beta)=\frac{\lambda}{\operatorname{LT}(\overline{\alpha})}\alpha - \frac{\lambda}{\operatorname{LT}(\overline{\beta})}\beta , \nonumber
\end{eqnarray}
where $\lambda$ is the least common monomial as $\lambda = \operatorname{lcm}(\operatorname{LT}(\overline{\alpha}),\operatorname{LT}(\overline{\beta}))$

Let $G$ be a subset of $R^m$, let $\alpha,\alpha' \in R^m$, we say that $\alpha$ is an {\it one-time $\mathfrak{s}$-reduced} to $\alpha'$ if there exist $\beta \in G$ and $b \in R$ satisfying:
\newpage
\begin{description}
\item[\qquad(a)] $\operatorname{LT}(\overline{b \beta})=t$ \qquad \qquad for a (certain) monomial in $\overline{\alpha}$
\item[\qquad(b)] $\mathfrak{s}(b \beta) \preceq \mathfrak{s}(\alpha)$
\item[\qquad(c)] $\alpha' = \alpha - b \beta$.
\end{description}
\noindent
At this time, we call $\beta$ a reducer.
We say that $\alpha$ is {\it $\mathfrak{s}$-reduced to $\alpha''$} with respect to $G$ if there exists a sequence $\alpha = \alpha^{(0)}$,$\alpha^{(1)}$,$\cdots$,$\alpha^{(l)} = \alpha''$ of $R^m$ such that $\alpha^{(i)}$ is one-time $\mathfrak{s}$-reduced to $\alpha^{(i+1)}$ with respect to $G$ for $i=0,1,\cdots,l-1$.
%Let $\alpha \in R^m$, $t \in \overline{\alpha}$, if there exist $\beta \in R^m$ and $b \in R$ satisfying (a) and (b), we say that $\alpha$ can be {\it $\mathfrak{s}$-reduced}.  
%Let $\alpha \in R^m$, $t \in \overline{\alpha}$, $\beta \in R^m$ and $b \in R$ satisfying (a) and (b), to get $\alpha - b \beta$ from $\alpha$ is called we {\it $\mathfrak{s}$-reduce} $\alpha$,  
%\begin{description}
%\item[(a)] $\operatorname{LT}(\overline{b \beta})=t$ 
%\item[(b)] $\mathfrak{s}(b \beta) \leq \mathfrak{s}(\alpha)$.
%\end{description}
A $\mathfrak{s}$-reduction is called a {\it singular $\mathfrak{s}$-reduction}, if there exists $c \in K$ such that 
$\mathfrak{s}(b \beta)= c \mathfrak{s}(\alpha)$ and 
otherwise it is called a {\it regular $\mathfrak{s}$-reduction}.
%In the algorithms, we are allowed to use only regular $\mathfrak{s}$-reduction.
If there exists $c \in K$ such that $\operatorname{LT}(b \overline{\beta})= c \operatorname{LT}(\overline{\alpha})$, the $\mathfrak{s}$-reduction is called a {\it one-time top $\mathfrak{s}$-reduction} and otherwise called a {\it one-time tail $\mathfrak{s}$-reduction}. 
{\bf Algorithm 1} shows that if an S-pair $\alpha$ can be one-time top $\mathfrak{s}$-reduced by $G$, it returns one-time top $\mathfrak{s}$-reduced $\alpha$ by $G$.
{\bf Algorithm 2} shows that if an S-pair $\alpha$ can be one-time tail $\mathfrak{s}$-reduced by $G$, it returns one-time tail $\mathfrak{s}$-reduced $\alpha$ by $G$.
If the $\alpha \in R^m$ cannot be regular top $\mathfrak{s}$-reduced, we say that $\alpha$ is {\it completely regular top $\mathfrak{s}$-reduced}.
If all the monomials in $\overline{\alpha}$ cannot be regular $\mathfrak{s}$-reduced, we call $\alpha$ is {\it completely regular full $\mathfrak{s}$-reduced}.
Below, to distinguish from $\mathfrak{s}$-reductions, the reductions used in Buchberger algorithms are called usual reductions.

\begin{table*}[t]
  \begin{center}
  %\caption{ }
  \label{newalg}
  \begin{tabular}{l} \hline \hline
    {\bf Algorithm 1} TOP\_REDUCE \qquad \qquad \qquad \qquad \qquad \qquad \qquad \qquad \qquad \\ \hline \hline%\qquad \qquad \qquad \qquad \qquad \qquad \qquad \qquad \qquad \\ \hline \hline
    {\bf input} : a finite subset $G \in R^m$, an S-pair $\alpha$ \\
    {\bf output} : an S-pair $\alpha$\\ 
    \hline
    %{\bf do}\\
    {\bf for} $\beta \in G$ {\bf do}\\
    \quad {\bf if} $\operatorname{LT}(\overline{\beta}) \mid \operatorname{LT}(\overline{\alpha})$ and $\mathfrak{s}(\alpha) \succ  \mathfrak{s}( \frac{\operatorname{LT}(\overline{\alpha})}{\operatorname{LT}(\overline{\beta}))} \cdot \beta)$ {\bf then} \\
    \quad \quad    $\alpha \leftarrow \alpha - \frac{\operatorname{LT}(\overline{\alpha})}{\operatorname{LT}(\overline{\beta}))} \cdot \beta$ \\
    \quad \quad   {\bf return} $\alpha$ \\ 
    \quad {\bf end if}  \\
    {\bf end for}  \\
    {\bf return} $\alpha$ \\ \hline 
    \end{tabular}
    \end{center}
    \begin{center}
    %\caption{ }
    \label{newalg}
    \begin{tabular}{l} \hline \hline
    {\bf Algorithm 2} TAIL\_REDUCE \qquad \qquad \qquad \qquad \qquad \qquad \qquad \qquad \qquad \\ \hline \hline%\qquad \qquad \qquad \qquad \qquad \qquad \qquad \qquad \qquad \\ \hline \hline
    {\bf input} : a finite subset $G \in R^m$, an S-pair $\alpha$ \\
    {\bf output} : an S-pair $\alpha$ \\ 
    \hline
    %{\bf do}\\
    {\bf for} $t \in \operatorname{T}(\overline{\alpha} - \operatorname{LT}(\overline{\alpha}))$ {\bf do} \qquad \qquad \qquad ($t$ is a monomial in $\overline{\alpha} - \operatorname{LT}(\overline{\alpha})$)\\
    \quad {\bf for}  $\beta \in G$ {\bf do}\\ 
    \quad \quad {\bf if} $\operatorname{LT}(\overline{\beta}) \mid t$ and $\mathfrak{s}(\alpha) \succ \mathfrak{s}(\frac{t}{\operatorname{LT}(\overline{\beta}))} \cdot \beta)$ {\bf then} \\
    \quad \quad \quad $\alpha \leftarrow \alpha - \frac{t}{\operatorname{LT}(\overline{\beta}))} \cdot \beta$ \\
    \quad \quad \quad {\bf return} $\alpha$ \\ 
    \quad \quad {\bf end if}  \\
    \quad {\bf end for}  \\
    {\bf end for}  \\
    {\bf return} $\alpha$ \\ \hline 
  \end{tabular}
  \end{center}
\end{table*}

A subset $G \subseteq R^m$ is {\it a signature Gr\"obner basis with respect to $\preceq$} if all $\alpha \in R^m$ are $\mathfrak{s}$-reduced to zero with respect to $G$.
The signature-based algorithms compute a signature Gr\"obner basis.
If $G$ is a signature Gr\"obner basis, then \{$\overline{g} \mid g \in G$\} is a Gr\"obner basis of an ideal of \{$\overline{g} \mid g \in G$\}.
The number of the elements of a signature Gr\"obner basis is always larger than or equal to that of minimal Gr\"obner basis.

%Let $I \subset R$ be an ideal, an subset $G$ is a signature Gr\"obner basis of $I$ with regard to $\leq$ if all $\alpha \in R^m$ $\mathfrak{s}$-reduce to zero with regard to G.

%reducedグレブナ基底とminimalグレブナ基底
%spair

%\begin{itemize}
% \item[(a)] $LT(\overline{b \beta})=t$ 
% \item[(b)] $\mathfrak{s}(b \beta) \leq \mathfrak{s}{\alpha}$
%\end{itemize}

%$I=\langle f_1, \dots, f_m \rangle$を$R$のイデアルとする．
%有限集合$G \in R^m$が加群の項$T \in R^m$未満のsignature Gr\"obner基底であるとは，$\mathfrak{s}(\alpha)<T$となる全ての$\alpha \in R^m$に関して$G$により$\mathfrak{s}$-reduceして0になるときをいう．
%また，$G$がsignature Gr\"obner基底であるとは，全ての$\alpha \in R^m$が$\mathfrak{s}$-reduceで0になるときをいう．

%グレブナ基底$G \in R$がminimalグレブナ基底であるとは，以下の条件を満たす$g \in G$がないときである．
%\begin{eqnarray}
%\exists g' \in G,g' \neq g,\operatorname{LT}(g) \mid \operatorname{LT}(g')　\nonumber
%\end{eqnarray}
%グレブナ基底$G \in R$がreducedグレブナ基底であるとは，以下の条件を満たす$g \in G$がないときである．
%\begin{eqnarray}
%\exists g' \in G,g' \neq g,T(g) \mid \operatorname{LT}(g')　\nonumber
%\end{eqnarray}

\begin{table*}[t]
  \begin{center}
  %\caption{ }
  \label{newalg}
  \begin{tabular}{l} \hline \hline
    {\bf Algorithm 3} ONLY-TOP\_REDUCE \qquad \qquad \qquad \\ \hline \hline%\qquad \qquad \qquad \qquad \qquad \qquad \qquad \qquad \qquad \\ \hline \hline
    {\bf input} : a finite subset $G \in R^m$, an S-pair $\alpha$ \\
    {\bf output} : an S-pair $\alpha$ which is completely regular top $\mathfrak{s}$-reduced by $G$ \\ 
    \hline
    %{\bf do}\\
    $\beta \leftarrow 0$ \\
    {\bf while} $\beta \neq \alpha  $ {\bf do} \\
    \quad $\beta \leftarrow \alpha$ \\
    \quad $\alpha \leftarrow \operatorname{TOP\_REDUCE}(G,\alpha)$ \\ 
    {\bf end while} \\
    {\bf return} $\alpha$ \\ \hline 
  \end{tabular}
  \end{center}
\end{table*}

\begin{table*}[]
  \begin{center}
  %\caption{ }
  \label{newalg}
  \begin{tabular}{l} \hline \hline
    {\bf Algorithm 4} FULL\_REDUCE \qquad \qquad \qquad \\ \hline \hline %\qquad \qquad \qquad \qquad \qquad \qquad \qquad \qquad \qquad \qquad \\ \hline \hline
    {\bf input} : a finite subset $G \in R^m$, an S-pair $\alpha$ \\
    {\bf output} : an S-pair $\alpha$ which is completely regular full $\mathfrak{s}$-reduced by $G$ \\ 
    \hline
    %{\bf do}\\
    $\beta \leftarrow 0$ \\
    {\bf while} $\beta \neq \alpha$ {\bf do} \\
    \quad$\beta \leftarrow \alpha$ \\
    \quad $\alpha \leftarrow \operatorname{TOP\_REDUCE}(G,\alpha)$ \\ 
    {\bf end while} \\
    $\beta \leftarrow 0$ \\
    {\bf while} $\beta \neq \alpha $ {\bf do} \\
    \quad$\beta \leftarrow \alpha$ \\
    \quad $\alpha \leftarrow \operatorname{TAIL\_REDUCE}(G,\alpha)$ \\ 
    {\bf end while} \\
    {\bf return} $\alpha$ \\ \hline 
  \end{tabular}
  \end{center}
\end{table*}

\begin{table*}[]
  \begin{center}
  %\caption{ }
  \label{newalg}
  \begin{tabular}{l} \hline \hline
    {\bf Algorithm 5} SELECTIVE-FULL\_REDUCE \qquad \qquad \qquad \\ \hline \hline%\qquad \qquad \qquad \qquad \qquad \qquad \qquad\\ \hline \hline
    {\bf input} : a finite subset $G \in R^m$, an S-pair $\alpha$ \\
    {\bf output} : an S-pair $\alpha$ which is completely regular selective-full $\mathfrak{s}$-reduced by $G$ \\ 
    \hline
    %{\bf do}\\
    $\beta \leftarrow 0$ \\
    {\bf while} $\beta \neq \alpha$ {\bf do} \\
    \quad$\beta \leftarrow \alpha$ \\
    \quad $\alpha \leftarrow \operatorname{TOP\_REDUCE}(G,\alpha)$ \\ 
    {\bf end while} \\
    {\bf for}  $\gamma \in G$ {\bf do}\\
    \quad {\bf if} $\operatorname{LT}(\overline{\gamma}) \mid \operatorname{LT}(\overline{\alpha})$ {\bf then} \\
    \quad \quad {\bf return} $\alpha$ \qquad \qquad \qquad \qquad (if $\alpha$ does not satisfy ${\fbox{{\bf SF}}}$, return $\alpha$)\\ 
    \quad {\bf end if}  \\
    {\bf end for}  \\
    $\beta \leftarrow 0$ \\
    {\bf while} $\beta \neq \alpha$ {\bf do} \\
    \quad $\beta \leftarrow \alpha$ \\
    \quad $\alpha \leftarrow \operatorname{TAIL\_REDUCE}(G,\alpha)$ \\ 
    {\bf end while} \\
    {\bf return} $\alpha$ \\ \hline 
  \end{tabular}
  \end{center}
\end{table*}

\section{Conventional $\mathfrak{s}$-reduction strategies}
%S多項式⇒Spair

In this section, we review the two strategies of $\mathfrak{s}$-reducing S-pairs mentioned in the paper\cite{Eder17}.
One is the only-top reduction strategy: 
after generating an S-pair, regular $\mathfrak{s}$-reduce leading monomials until the leading monomial cannot be regular $\mathfrak{s}$-reduced. 
The algorithm is represented as {\bf Algorithm 3}.
By this procedure, the S-pair is completely regular top $\mathfrak{s}$-reduced. 
The other is the full reduction strategy: 
after generating S-pairs, regular $\mathfrak{s}$-reduce the monomials included in the S-pairs.
The algorithm is represented as {\bf Algorithm 4}.
At first, it execute top $\mathfrak{s}$-reduction, then if the S-pair is completely regular top $\mathfrak{s}$-reduced, execute regular tail $\mathfrak{s}$-reduction. 
By this procedure, the S-pair is completely regular full $\mathfrak{s}$-reduced.

Each strategy has advantages and disadvantages.
When we choose the only-top reduction strategy, it is expected that times of $\mathfrak{s}$-reduction is fewer because we regular $\mathfrak{s}$-reduce the only top monomials.
However, since the terms of the polynomials used to regular $\mathfrak{s}$-reduce the S-pairs are large with respect to the fixed monomial order, there is a possibility that the number of times of $\mathfrak{s}$-reduction may increase even if only the regular $\mathfrak{s}$-reduction of the leading term is performed.
On the other hand, when we choose the full reduction strategy, we regular $\mathfrak{s}$-reduce all terms included in the S-pair, so the terms included in the S-pairs is relatively small with respect to the fixed monomial order.
Also, times of interreductions for computing the reduced Gr\"obner basis becomes few because regular tail $\mathfrak{s}$-reductions has executed in advance.
However, regular tail $\mathfrak{s}$-reductions are restricted reductions, so it cannot reduce terms sufficiently in comparison with usual reductions.
Moreover, number of elements of a signature Gr\"obner basis tend to be much larger than that of minimal Gr\"obner basis.
So, the number of S-pairs that we need to completely regular full $\mathfrak{s}$-reduce is also much larger.

\section{Our $\mathfrak{s}$-reduction strategy}
Consider the case where we compute the reduced Gr\"obner basis after a signature Gr\"obner basis is computed.
The signature-based algorithms compute the signature Gr\"obner basis which is larger than the minimnal Gr\"obner basis.
Therefore, first, compute a minimal Gr\"obner basis from the found signature Gr\"obner basis.
The method is to remove $\alpha \in G$ satisfying the following condition from the found signature Gr\"obner basis: 
There exists $\alpha' \in G,\operatorname{LT}(\overline{\alpha}) \mid \operatorname{LT}(\overline{\alpha'})$.
Then, we obtain a minimal Gr\"obner basis.
By interreducing the found minimal Gr\"obner basis, the reduced Gr\"obner basis is obtained.

Here we consider the relation between the full reduction strategy and reduced Gr\"obner basis.
The full reduction strategy can be seen as a strategy to decrease the number of times of interreduction.
In that sense, there is no need to tail reduction for S-pairs that will be removed at the step of computing a minimal Gr\"obner basis.
An algorithm based on this idea to $\mathfrak{s}$-reduce an S-pair is {\bf Algorithm 5}.
In this algorithm, first, regular top $\mathfrak{s}$-reduce the S-pair $\alpha$ until the S-pair becomes completely regular top $\mathfrak{s}$-reduced.
Then, perform regular tail $\mathfrak{s}$-reduction only when the following condition is satisfied: 
\begin{equation*}
\text{for all }\alpha' \in G,\operatorname{LT}(\overline{\alpha'}) \nmid \operatorname{LT}(\overline{\alpha})   \eqno {\fbox{{\bf SF}}} 
\end{equation*}
We call this strategy the {\it selective-full reduction}.
The output of the {\bf Algorithm 5} denotes a {\it completely regular selective-full reduced} S-pair.

Following shows that the selective-full strategy is reasonable.
\begin{description}
\item[(1)] Let $\alpha$ be an S-pair which does not satisfy {\fbox{{\bf SF}}}.
We can foresee that $\alpha$ will be removed when we compute a minimal Gr\"obner basis.
If we choose the selective-full strategy, we do not regular tail $\mathfrak{s}$-reduce $\alpha$ that is finally discarded.
Therefore, it is expected that number of times of $\mathfrak{s}$-reductions by the selective-full reduction strategy is smaller that by full reduction strategy.
\item[(2)] Consider the case that a signature Gr\"obner basis has been computed, and then, we obtain an minimal Gr\"obner basis.
If we choose the selective-full strategy, all elements in the minimal Gr\"obner basis were completely regular full $\mathfrak{s}$-reduced.
Therefore, it is expected that number of times of interreductions by the selective-full reduction strategy is much smaller than that by only-top reduction strategy.
\item[(3)] Let $\alpha \in G$ be a possible reducer for a certain S-pair, and $\alpha$ was not completely regular full $\mathfrak{s}$-reduced, that means that $\alpha$ did not satisfy {\fbox{{\bf SF}}}.
Then, there exists $\alpha'$ such that $\operatorname{LT}(\overline{\alpha'}) \mid \operatorname{LT}(\overline{\alpha})$ and $\alpha'$ was completely regular full $\mathfrak{s}$-reduced or an input module of the algorithm.
Because, in the above situation, there exists $\alpha' \in G$ such that $\operatorname{LT}(\overline{\alpha'}) \mid \operatorname{LT}(\overline{\alpha})$ and $\operatorname{LT}(\overline{\beta}) \nmid \operatorname{LT}(\overline{\alpha'})$ for all $\beta \in G$.
Then, $\alpha'$ was generated from a certain S-pair that satisfy {\fbox{{\bf SF}}}, or is an input module of the algorithm.
So, if the S-pair can be regular $\mathfrak{s}$-reduced, a reducer which is regular full $\mathfrak{s}$-reduced is possible to be selected.
Therefore, to some extent, number of times of $\mathfrak{s}$-reduction is expected to be less number compared to the only-top reduction strategy.
\end{description}

\section{Results}
In this section, we evaluate the proposed $\mathfrak{s}$-reduction strategy, selective-full reduction strategy, using the well-known benchmark for Gr\"obner basis.
The implementation is done by C for counting $\mathfrak{s}$-reductions, usual reductions and multiplications.
The benchmark was carried out in each of homogeneous ideals and inhomogeneous ideals.
We compared three strategies, only-top reduce, full reduce, and selective-full reduce.
We refer to \cite{Eder17} for the concept of the signature-based algorithm and the words used below.
All systems are computed over a field of characteristic 32003, with graded reverse lexicographical monomial order.
For a module order, we used the POT order which is used in the original F5.
For finding the syzygy modules, we used signatures which are zero reduced.
Therefore, like F5, all algorithms proceeds incrementally.
Like F5C, the reduced Gr\"obner basis was found at each incremental steps.
For a rewrite order, we used each ADD and RAT.
\begin{remark*}
In the {\bf Algorithm 2}, there is no restriction to choose a monomial in $\overline{\alpha} - \operatorname{LT}(\overline{\alpha})$.
In the experiment, we regular $\mathfrak{s}$-reduced a monomial in large monomial order.
By doing so, monomials which is larger than a regular $\mathfrak{s}$-reduced monomial do not vary.
\end{remark*}

The results are shown in table 1,2,3,4,5,6. 
In table 1,3, the numbers of sum of one-time $\mathfrak{s}$-reductions and usual reductions to compute a signature Gr\"obner basis(SGB:ALL), 
among them the numbers of one-time $\mathfrak{s}$-reductions(SGB:S-RED) 
and the numbers of sum of one-time $\mathfrak{s}$-reductions and usual reductions to compute the reduced Gr\"obner basis(RGB:ALL) 
are shown.
In table 2,4, the numbers of times of multiplications processed in above computation are shown.
In table 5,6, the numbers of generated S-pairs which satisfy {\fbox{{\bf SF}}} and does not satisfy {\fbox{{\bf SF}}} are shown.

For computing the reduced Gr\"obner basis, the selective-full reduction strategy processes the least times of sum of one-time $\mathfrak{s}$-reductions and usual reductions of the three strategies.
Also, it processes less times of multiplications except for little disadvantage to full reduction strategy at some benchmarks.
Therefore, the selective-full reduction strategy is efficient for the reduced Gr\"obner basis.

For computing a signature Gr\"obner basis, when comparing the only-top reduction strategy with the selective-full strategy, the selective-full strategy is better or worse for some benchmarks.
When comparing the full reduction strategy with the selective-full strategy, the selective-full reduction strategy is better or equivalent.
From table 5,6, the more effective the selective-full reduction strategy is, the more number the  difference between the reduced Gr\"obner basis and a signature Gr\"obner basis is. 
The only-top reduction strategy is weak in Random(10,2,2) and Random(11,2,2)(both homogeneous and inhomogeneous) and the full reduction strategy is weak in katstura-11(both homogeneous and inhomogeneous).
Although, the selective-full strategy is not weak in the above three.
Besides, the selective-full strategy is not the worst strategy in the three strategies in the case of all benchmarks in this paper.

%With the selective-full strategy, the numbers of (SGB:ALL) were the smallest, or equivalent to the fewest except for noon-8 and 9.
%This indicates that the selective-full strategy has an advantage in most cases.
%In addition, the numbers of times of $\mathfrak{s}$-reduction which is more troublesome than usual reduction is also small in most cases.
%Especially, the selective-full strategy can compute the reduced Gr\"obner basis with the less number of sum of $\mathfrak{s}$-reductions and usual reductions than the other two.

\section{Conclusion}
We propose a new strategy for $\mathfrak{s}$-reduction and evaluated it with benchmarks.
For computing the reduced Gr\"obner basis, the selective-full reduction strategy is more efficient comparing with convetional $\mathfrak{s}$-reduction strategies.
For computing a signature Gr\"obner basis, the selective-full reduction strategy is better or equivalent to the full reduction strategy.
It depends on the system that which the selective-full reduction strategy or the only-top reduction strategy is better.
Although, the selective-full strategy is not the worst strategy in the three strategies in the case of all benchmarks in this paper.
In the future, based on the result from this paper, we develop the signature based algorithms for computing Gr\"obner basis more efficiently.

\begin{table}[p]
  \begin{center}
% \normalsize
  \caption{The numbers of times of reductions (homogeneous)} 
  \begin{tabular}{|l|c|r|r||r||r|r||r|} \hline
    \multicolumn{1}{|l|}{}
      & \multicolumn{1}{l|}{}
      & \multicolumn{3}{|c||}{ADD} 
      & \multicolumn{3}{c|}{RAT} \\ \cline{3-8}
    \multicolumn{1}{|l|}{benchmark}
      & \multicolumn{1}{c|}{} 
      & \multicolumn{2}{c||}{SGB} 
      & \multicolumn{1}{c||}{RGB}
      & \multicolumn{2}{c||}{SGB} 
      & \multicolumn{1}{c|}{RGB}\\ \cline{3-8}
    \multicolumn{1}{|l|}{}
      & \multicolumn{1}{c|}{} 
      & \multicolumn{1}{c|}{ALL} 
      & \multicolumn{1}{c||}{$\mathfrak{s}$-RED} 
      & \multicolumn{1}{c||}{ALL} 
      & \multicolumn{1}{c|}{ALL} 
      & \multicolumn{1}{c||}{$\mathfrak{s}$-RED} 
      & \multicolumn{1}{c|}{ALL} \\ \hline \hline
                   & only-top & $2^{17.380
}$ & $2^{17.118
}$ & $2^{17.528
}$ & $2^{17.099
}$ & $2^{16.740
}$ & $2^{17.284
}$ \\
    cyclic-7       & full     & $2^{16.954
}$ & $2^{16.936
}$ & $2^{17.035
}$ & $2^{16.442
}$ & $2^{16.416
}$ & $2^{16.557
}$ \\
                   & selective& $2^{16.699
}$ & $2^{16.677
}$ & $2^{16.795
}$ & $2^{16.344
}$ & $2^{16.316
}$ & $2^{16.466
}$ \\ \hline
                   & only-top & $2^{22.788
}$ & $2^{22.484
}$ & $2^{22.837
}$ & $2^{21.983
}$ & $2^{21.382
}$ & $2^{22.075
}$ \\
    cyclic-8       & full     & $2^{23.319
}$ & $2^{23.298
}$ & $2^{23.333
}$ & $2^{22.041
}$ & $2^{21.990
}$ & $2^{22.076
}$ \\
                   & selective& $2^{22.336
}$ & $2^{22.295
}$ & $2^{22.365
}$ & $2^{21.280
}$ & $2^{21.192
}$ & $2^{21.339
}$ \\ \hline
                   & only-top & $2^{19.026
}$ & $2^{18.902
}$ & $2^{19.435
}$ & $2^{18.781
}$ & $2^{18.632
}$ & $2^{19.255
}$ \\
    eco-10         & full     & $2^{20.314
}$ & $2^{20.302
}$ & $2^{20.350
}$ & $2^{19.013
}$ & $2^{18.983
}$ & $2^{19.101
}$ \\
                   & selective& $2^{18.852
}$ & $2^{18.819
}$ & $2^{18.950
}$ & $2^{18.741
}$ & $2^{18.704
}$ & $2^{18.846
}$ \\ \hline
                   & only-top & $2^{21.541
}$ & $2^{21.421
}$ & $2^{21.950
}$ & $2^{21.166
}$ & $2^{21.008
}$ & $2^{21.679
}$ \\
    eco-11         & full     & $2^{23.739
}$ & $2^{23.734
}$ & $2^{23.755
}$ & $2^{21.486
}$ & $2^{21.465
}$ & $2^{21.563
}$ \\
                   & selective& $2^{21.401
}$ & $2^{21.378
}$ & $2^{21.482
}$ & $2^{21.166
}$ & $2^{21.139
}$ & $2^{21.261
}$ \\ \hline
                   & only-top & $2^{9.852
}$ & $2^{9.718
}$ & $2^{10.723
}$ & $2^{9.647
}$ & $2^{9.492
}$ & $2^{10.530
}$ \\
    f-633          & full     & $2^{9.990
}$ & $2^{9.956
}$ & $2^{10.007
}$ & $2^{9.533
}$ & $2^{9.486
}$ & $2^{9.557
}$ \\
                   & selective& $2^{9.635
}$ & $2^{9.591
}$ & $2^{9.656
}$ & $2^{9.496
}$ & $2^{9.447
}$ & $2^{9.520
}$ \\ \hline
                   & only-top & $2^{16.942
}$ & $2^{16.752
}$ & $2^{17.074
}$ & $2^{16.598
}$ & $2^{16.348
}$ & $2^{16.757
}$ \\
    f-744          & full     & $2^{17.398
}$ & $2^{17.393
}$ & $2^{17.414
}$ & $2^{16.435
}$ & $2^{16.426
}$ & $2^{16.465
}$ \\
                   & selective& $2^{16.801
}$ & $2^{16.795
}$ & $2^{16.825
}$ & $2^{16.340
}$ & $2^{16.331
}$ & $2^{16.373
}$ \\ \hline
                   & only-top & $2^{21.797
}$ & $2^{18.644
}$ & $2^{22.331
}$ & $2^{21.747
}$ & $2^{18.356
}$ & $2^{22.257
}$ \\
    katsura-11     & full     & $2^{23.709
}$ & $2^{23.700
}$ & $2^{23.713
}$ & $2^{23.515
}$ & $2^{23.504
}$ & $2^{23.520
}$ \\
                   & selective& $2^{21.600
}$ & $2^{21.560
}$ & $2^{21.618
}$ & $2^{21.594
}$ & $2^{21.553
}$ & $2^{21.612
}$ \\ \hline
                   & only-top & $2^{15.849
}$ & $2^{15.847
}$ & $2^{20.145
}$ & $2^{14.541
}$ & $2^{14.537
}$ & $2^{19.993
}$ \\
    noon-8         & full     & $2^{18.103
}$ & $2^{18.103
}$ & $2^{18.109
}$ & $2^{17.940
}$ & $2^{17.940
}$ & $2^{17.948
}$ \\
                   & selective& $2^{18.024
}$ & $2^{18.024
}$ & $2^{18.031
}$ & $2^{17.865
}$ & $2^{17.865
}$ & $2^{17.873
}$ \\ \hline
                   & only-top & $2^{18.401
}$ & $2^{18.400
}$ & $2^{23.045
}$ & $2^{16.756
}$ & $2^{16.755
}$ & $2^{22.881
}$ \\
    noon-9         & full     & $2^{20.685
}$ & $2^{20.685
}$ & $2^{20.690
}$ & $2^{20.515
}$ & $2^{20.515
}$ & $2^{20.521
}$ \\
                   & selective& $2^{20.603
}$ & $2^{20.603
}$ & $2^{20.608
}$ & $2^{20.445
}$ & $2^{20.445
}$ & $2^{20.451
}$ \\ \hline
                   & only-top & $2^{19.452
}$ & $2^{18.004
}$ & $2^{19.575
}$ & $2^{19.454
}$ & $2^{18.009
}$ & $2^{19.577
}$ \\
    HRandom(10,2,2)& full     & $2^{17.659
}$ & $2^{17.616
}$ & $2^{17.761
}$ & $2^{17.660
}$ & $2^{17.617
}$ & $2^{17.762
}$ \\
                   & selective& $2^{17.655
}$ & $2^{17.611
}$ & $2^{17.757
}$ & $2^{17.656
}$ & $2^{17.613
}$ & $2^{17.758
}$ \\ \hline
                   & only-top & $2^{21.620
}$ & $2^{20.097
}$ & $2^{21.727
}$ & $2^{21.621
}$ & $2^{20.101
}$ & $2^{21.729
}$ \\
    HRandom(11,2,2)& full     & $2^{19.612
}$ & $2^{19.571
}$ & $2^{19.690
}$ & $2^{19.612
}$ & $2^{19.571
}$ & $2^{19.691
}$ \\
                   & selective& $2^{19.596
}$ & $2^{19.555
}$ & $2^{19.676
}$ & $2^{19.598
}$ & $2^{19.556
}$ & $2^{19.677
}$ \\ \hline
  \end{tabular}
  \end{center}
\end{table}

\begin{table}[p]
  \begin{center}
%\footnotesize
  \caption{The numbers of times of multiplications (homogeneous)} 
  \begin{tabular}{|l|c|r|r||r||r|r||r|} \hline
    \multicolumn{1}{|l|}{}
      & \multicolumn{1}{l|}{}
      & \multicolumn{3}{|c||}{ADD} 
      & \multicolumn{3}{c|}{RAT} \\ \cline{3-8}
    \multicolumn{1}{|l|}{benchmark}
      & \multicolumn{1}{c|}{} 
      & \multicolumn{2}{c||}{SGB} 
      & \multicolumn{1}{c||}{RGB}
      & \multicolumn{2}{c||}{SGB} 
      & \multicolumn{1}{c|}{RGB}\\ \cline{3-8}
    \multicolumn{1}{|l|}{}
      & \multicolumn{1}{c|}{} 
      & \multicolumn{1}{c|}{ALL} 
      & \multicolumn{1}{c||}{$\mathfrak{s}$-RED} 
      & \multicolumn{1}{c||}{ALL} 
      & \multicolumn{1}{c|}{ALL} 
      & \multicolumn{1}{c||}{$\mathfrak{s}$-RED} 
      & \multicolumn{1}{c|}{ALL} \\ \hline \hline
                   & only-top & $2^{24.401
}$ & $2^{24.275
}$ & $2^{24.523
}$ & $2^{23.994
}$ & $2^{23.820
}$ & $2^{24.155
}$ \\
    cyclic-7       & full     & $2^{24.305
}$ & $2^{24.286
}$ & $2^{24.400
}$ & $2^{23.758
}$ & $2^{23.730
}$ & $2^{23.895
}$ \\
                   & selective& $2^{24.057
}$ & $2^{24.034
}$ & $2^{24.169
}$ & $2^{23.616
}$ & $2^{23.584
}$ & $2^{23.766
}$ \\ \hline
                   & only-top & $2^{31.140
}$ & $2^{30.971
}$ & $2^{31.181
}$ & $2^{30.132
}$ & $2^{29.769
}$ & $2^{30.214
}$ \\
    cyclic-8       & full     & $2^{31.978
}$ & $2^{31.959
}$ & $2^{31.995
}$ & $2^{30.698
}$ & $2^{30.650
}$ & $2^{30.739
}$ \\
                   & selective& $2^{30.861
}$ & $2^{30.818
}$ & $2^{30.898
}$ & $2^{29.789
}$ & $2^{29.696
}$ & $2^{29.865
}$ \\ \hline
                   & only-top & $2^{24.460
}$ & $2^{24.340
}$ & $2^{24.839
}$ & $2^{24.291
}$ & $2^{24.155
}$ & $2^{24.713
}$ \\
    eco-10         & full     & $2^{26.007
}$ & $2^{25.996
}$ & $2^{26.043
}$ & $2^{24.708
}$ & $2^{24.681
}$ & $2^{24.795
}$ \\
                   & selective& $2^{24.513
}$ & $2^{24.481
}$ & $2^{24.612
}$ & $2^{24.409
}$ & $2^{24.375
}$ & $2^{24.515
}$ \\ \hline
                   & only-top & $2^{27.628
}$ & $2^{27.505
}$ & $2^{27.992
}$ & $2^{27.339
}$ & $2^{27.187
}$ & $2^{27.775
}$ \\
    eco-11         & full     & $2^{29.978
}$ & $2^{29.974
}$ & $2^{29.996
}$ & $2^{27.812
}$ & $2^{27.791
}$ & $2^{27.889
}$ \\
                   & selective& $2^{27.711
}$ & $2^{27.689
}$ & $2^{27.793
}$ & $2^{27.484
}$ & $2^{27.458
}$ & $2^{27.580
}$ \\ \hline
                   & only-top & $2^{12.429
}$ & $2^{12.302
}$ & $2^{13.305
}$ & $2^{12.232
}$ & $2^{12.086
}$ & $2^{13.105
}$ \\
    f-633          & full     & $2^{12.687
}$ & $2^{12.649
}$ & $2^{12.702
}$ & $2^{12.300
}$ & $2^{12.251
}$ & $2^{12.321
}$ \\
                   & selective& $2^{12.356
}$ & $2^{12.309
}$ & $2^{12.376
}$ & $2^{12.258
}$ & $2^{12.207
}$ & $2^{12.279
}$ \\ \hline
                   & only-top & $2^{21.553
}$ & $2^{21.438
}$ & $2^{21.701
}$ & $2^{21.258
}$ & $2^{21.111
}$ & $2^{21.433
}$ \\
    f-744          & full     & $2^{22.201
}$ & $2^{22.198
}$ & $2^{22.223
}$ & $2^{21.274
}$ & $2^{21.267
}$ & $2^{21.315
}$ \\
                   & selective& $2^{21.554
}$ & $2^{21.549
}$ & $2^{21.588
}$ & $2^{21.152
}$ & $2^{21.145
}$ & $2^{21.196
}$ \\ \hline
                   & only-top & $2^{29.633
}$ & $2^{27.907
}$ & $2^{30.009
}$ & $2^{29.548
}$ & $2^{27.635
}$ & $2^{29.926
}$ \\
    katsura-11     & full     & $2^{32.170
}$ & $2^{32.162
}$ & $2^{32.175
}$ & $2^{31.969
}$ & $2^{31.961
}$ & $2^{31.976
}$ \\
                   & selective& $2^{29.932
}$ & $2^{29.897
}$ & $2^{29.958
}$ & $2^{29.907
}$ & $2^{29.871
}$ & $2^{29.934
}$ \\ \hline
                   & only-top & $2^{20.402
}$ & $2^{20.401
}$ & $2^{23.998
}$ & $2^{19.802
}$ & $2^{19.801
}$ & $2^{23.906
}$ \\
    noon-8         & full     & $2^{22.249
}$ & $2^{22.249
}$ & $2^{22.388
}$ & $2^{22.128
}$ & $2^{22.128
}$ & $2^{22.279
}$ \\
                   & selective& $2^{22.169
}$ & $2^{22.169
}$ & $2^{22.316
}$ & $2^{22.059
}$ & $2^{22.059
}$ & $2^{22.218
}$ \\ \hline
                   & only-top & $2^{23.198
}$ & $2^{23.198
}$ & $2^{27.216
}$ & $2^{22.394
}$ & $2^{22.393
}$ & $2^{27.110
}$ \\
    noon-9         & full     & $2^{25.156
}$ & $2^{25.156
}$ & $2^{25.332
}$ & $2^{25.029
}$ & $2^{25.029
}$ & $2^{25.220
}$ \\
                   & selective& $2^{25.076
}$ & $2^{25.076
}$ & $2^{25.261
}$ & $2^{24.969
}$ & $2^{24.969
}$ & $2^{25.167
}$ \\ \hline
                   & only-top & $2^{27.035
}$ & $2^{26.085
}$ & $2^{27.103
}$ & $2^{27.036
}$ & $2^{26.087
}$ & $2^{27.105
}$ \\
    Random(10,2,2) & full     & $2^{25.936
}$ & $2^{25.872
}$ & $2^{25.984
}$ & $2^{25.937
}$ & $2^{25.874
}$ & $2^{25.985
}$ \\
                   & selective& $2^{26.038
}$ & $2^{25.979
}$ & $2^{26.083
}$ & $2^{26.039
}$ & $2^{25.980
}$ & $2^{26.084
}$ \\ \hline
                   & only-top & $2^{29.967
}$ & $2^{29.024
}$ & $2^{30.022
}$ & $2^{29.968
}$ & $2^{29.025
}$ & $2^{30.023
}$ \\
    Random(11,2,2) & full     & $2^{28.831
}$ & $2^{28.770
}$ & $2^{28.867
}$ & $2^{28.831
}$ & $2^{28.770
}$ & $2^{28.867
}$ \\
                   & selective& $2^{28.937
}$ & $2^{28.880
}$ & $2^{28.970
}$ & $2^{28.938
}$ & $2^{28.882
}$ & $2^{28.972
}$ \\ \hline
  \end{tabular}
  \end{center}
\end{table}

%                   & only-top & 0 & 1 & 2 & 0 & 1 & 2 \\
%    HRandom(7,2,4) & full     & 0 & 1 & 2 & 0 & 1 & 2 \\
%                   & selective& 0 & 1 & 2 & 0 & 1 & 2 \\ \hline
%                   & only-top & 0 & 1 & 2 & 0 & 1 & 2 \\
%    HRandom(8,2,4) & full     & 0 & 1 & 2 & 0 & 1 & 2 \\
%                   & selective& 0 & 1 & 2 & 0 & 1 & 2 \\ \hline

\begin{table}[p]
  \begin{center}
%\footnotesize
  \caption{The numbers of times of reductions (inhomogeneous)} 
  \begin{tabular}{|l|c|r|r||r||r|r||r|} \hline
    \multicolumn{1}{|l|}{}
      & \multicolumn{1}{l|}{}
      & \multicolumn{3}{|c||}{ADD} 
      & \multicolumn{3}{c|}{RAT} \\ \cline{3-8}
    \multicolumn{1}{|l|}{benchmark}
      & \multicolumn{1}{c|}{} 
      & \multicolumn{2}{c||}{SGB} 
      & \multicolumn{1}{c||}{RGB}
      & \multicolumn{2}{c||}{SGB} 
      & \multicolumn{1}{c|}{RGB}\\ \cline{3-8}
    \multicolumn{1}{|l|}{}
      & \multicolumn{1}{c|}{} 
      & \multicolumn{1}{c|}{ALL} 
      & \multicolumn{1}{c||}{$\mathfrak{s}$-RED} 
      & \multicolumn{1}{c||}{ALL} 
      & \multicolumn{1}{c|}{ALL} 
      & \multicolumn{1}{c||}{$\mathfrak{s}$-RED} 
      & \multicolumn{1}{c|}{ALL} \\ \hline \hline
                   & only-top & $2^{17.380
}$ & $2^{17.118
}$ & $2^{17.433
}$ & $2^{17.099
}$ & $2^{16.740
}$ & $2^{17.163
}$ \\
    cyclic-7       & full     & $2^{16.954
}$ & $2^{16.936
}$ & $2^{16.979
}$ & $2^{16.442
}$ & $2^{16.416
}$ & $2^{16.478
}$ \\
                   & selective& $2^{16.699
}$ & $2^{16.677
}$ & $2^{16.729
}$ & $2^{16.344
}$ & $2^{16.316
}$ & $2^{16.382
}$ \\ \hline
                   & only-top & $2^{22.788
}$ & $2^{22.484
}$ & $2^{22.793
}$ & $2^{21.983
}$ & $2^{21.382
}$ & $2^{21.994
}$ \\
    cyclic-8       & full     & $2^{23.319
}$ & $2^{23.298
}$ & $2^{23.320
}$ & $2^{22.041
}$ & $2^{21.990
}$ & $2^{22.046
}$ \\
                   & selective& $2^{22.355
}$ & $2^{22.314
}$ & $2^{22.358
}$ & $2^{21.288
}$ & $2^{21.200
}$ & $2^{21.295
}$ \\ \hline
                   & only-top & $2^{17.942
}$ & $2^{17.650
}$ & $2^{18.422
}$ & $2^{16.892
}$ & $2^{16.252
}$ & $2^{17.741
}$ \\
    eco-10         & full     & $2^{20.878
}$ & $2^{20.868
}$ & $2^{20.888
}$ & $2^{18.283
}$ & $2^{18.223
}$ & $2^{18.346
}$ \\
                   & selective& $2^{18.031
}$ & $2^{17.960
}$ & $2^{18.106
}$ & $2^{17.528
}$ & $2^{17.427
}$ & $2^{17.634
}$ \\ \hline
                   & only-top & $2^{20.637
}$ & $2^{20.369
}$ & $2^{21.048
}$ & $2^{19.125
}$ & $2^{18.247
}$ & $2^{20.064
}$ \\
    eco-11         & full     & $2^{24.492
}$ & $2^{24.489
}$ & $2^{24.495
}$ & $2^{20.790
}$ & $2^{20.742
}$ & $2^{20.825
}$ \\
                   & selective& $2^{20.710
}$ & $2^{20.659
}$ & $2^{20.747
}$ & $2^{19.890
}$ & $2^{19.797
}$ & $2^{19.954
}$ \\ \hline
                   & only-top & $2^{9.716
}$ & $2^{9.568
}$ & $2^{10.654
}$ & $2^{9.502
}$ & $2^{9.329
}$ & $2^{10.460
}$ \\
    f-633          & full     & $2^{9.950
}$ & $2^{9.914
}$ & $2^{9.979
}$ & $2^{9.474
}$ & $2^{9.424
}$ & $2^{9.514
}$ \\
                   & selective& $2^{9.583
}$ & $2^{9.537
}$ & $2^{9.620
}$ & $2^{9.435
}$ & $2^{9.384
}$ & $2^{9.476
}$ \\ \hline
                   & only-top & $2^{16.039
}$ & $2^{15.449
}$ & $2^{16.148
}$ & $2^{15.398
}$ & $2^{14.364
}$ & $2^{15.561
}$ \\
    f-744          & full     & $2^{16.286
}$ & $2^{16.278
}$ & $2^{16.313
}$ & $2^{15.424
}$ & $2^{15.409
}$ & $2^{15.472
}$ \\
                   & selective& $2^{15.560
}$ & $2^{15.546
}$ & $2^{15.604
}$ & $2^{14.458
}$ & $2^{14.427
}$ & $2^{14.550
}$ \\ \hline
                   & only-top & $2^{21.797
}$ & $2^{18.644
}$ & $2^{22.331
}$ & $2^{21.747
}$ & $2^{18.356
}$ & $2^{22.257
}$ \\
    katsura-11     & full     & $2^{23.709
}$ & $2^{23.700
}$ & $2^{23.713
}$ & $2^{23.515
}$ & $2^{23.504
}$ & $2^{23.520
}$ \\
                   & selective& $2^{21.600
}$ & $2^{21.560
}$ & $2^{21.618
}$ & $2^{21.594
}$ & $2^{21.553
}$ & $2^{21.612
}$ \\ \hline
                   & only-top & $2^{15.849
}$ & $2^{15.847
}$ & $2^{20.145
}$ & $2^{14.541
}$ & $2^{14.537
}$ & $2^{19.993
}$ \\
    noon-8         & full     & $2^{17.940
}$ & $2^{17.940
}$ & $2^{17.948
}$ & $2^{18.011
}$ & $2^{18.011
}$ & $2^{18.018
}$ \\
                   & selective& $2^{18.024
}$ & $2^{18.024
}$ & $2^{18.031
}$ & $2^{17.865
}$ & $2^{17.865
}$ & $2^{17.873
}$ \\ \hline
                   & only-top & $2^{18.401
}$ & $2^{18.400
}$ & $2^{23.045
}$ & $2^{16.756
}$ & $2^{16.755
}$ & $2^{22.881
}$ \\
    noon-9         & full     & $2^{20.515
}$ & $2^{20.515
}$ & $2^{20.521
}$ & $2^{20.584
}$ & $2^{20.584
}$ & $2^{20.589
}$ \\
                   & selective& $2^{20.603
}$ & $2^{20.603
}$ & $2^{20.608
}$ & $2^{20.445
}$ & $2^{20.445
}$ & $2^{20.451
}$ \\ \hline
                   & only-top & $2^{19.452
}$ & $2^{18.004
}$ & $2^{19.575
}$ & $2^{19.454
}$ & $2^{18.009
}$ & $2^{19.577
}$ \\
    Random(10,2,2) & full     & $2^{17.659
}$ & $2^{17.616
}$ & $2^{17.761
}$ & $2^{17.660
}$ & $2^{17.617
}$ & $2^{17.762
}$ \\
                   & selective& $2^{17.655
}$ & $2^{17.611
}$ & $2^{17.757
}$ & $2^{17.656
}$ & $2^{17.613
}$ & $2^{17.758
}$ \\ \hline
                   & only-top & $2^{21.620
}$ & $2^{20.097
}$ & $2^{21.727
}$ & $2^{21.621
}$ & $2^{20.101
}$ & $2^{21.729
}$ \\
    Random(11,2,2) & full     & $2^{19.612
}$ & $2^{19.571
}$ & $2^{19.690
}$ & $2^{19.612
}$ & $2^{19.571
}$ & $2^{19.691
}$ \\
                   & selective& $2^{19.596
}$ & $2^{19.555
}$ & $2^{19.676
}$ & $2^{19.598
}$ & $2^{19.556
}$ & $2^{19.677
}$ \\ \hline
  \end{tabular}
  \end{center}
\end{table}

\begin{table}[p]
  \begin{center}
%\footnotesize
  \caption{The numbers of times of multiplications (inhomogeneous)} 
  \begin{tabular}{|l|c|r|r||r||r|r||r|} \hline
    \multicolumn{1}{|l|}{}
      & \multicolumn{1}{l|}{}
      & \multicolumn{3}{|c||}{ADD} 
      & \multicolumn{3}{c|}{RAT} \\ \cline{3-8}
    \multicolumn{1}{|l|}{benchmark}
      & \multicolumn{1}{c|}{} 
      & \multicolumn{2}{c||}{SGB} 
      & \multicolumn{1}{c||}{RGB}
      & \multicolumn{2}{c||}{SGB} 
      & \multicolumn{1}{c|}{RGB}\\ \cline{3-8}
    \multicolumn{1}{|l|}{}
      & \multicolumn{1}{c|}{} 
      & \multicolumn{1}{c|}{ALL} 
      & \multicolumn{1}{c||}{$\mathfrak{s}$-RED} 
      & \multicolumn{1}{c||}{ALL} 
      & \multicolumn{1}{c|}{ALL} 
      & \multicolumn{1}{c||}{$\mathfrak{s}$-RED} 
      & \multicolumn{1}{c|}{ALL} \\ \hline \hline
                   & only-top & $2^{24.401
}$ & $2^{24.275
}$ & $2^{24.438
}$ & $2^{23.994
}$ & $2^{23.820
}$ & $2^{24.042
}$ \\
    cyclic-7       & full     & $2^{24.305
}$ & $2^{24.286
}$ & $2^{24.326
}$ & $2^{23.758
}$ & $2^{23.730
}$ & $2^{23.788
}$ \\
                   & selective& $2^{24.057
}$ & $2^{24.034
}$ & $2^{24.081
}$ & $2^{23.616
}$ & $2^{23.584
}$ & $2^{23.649
}$ \\ \hline
                   & only-top & $2^{31.140
}$ & $2^{30.971
}$ & $2^{31.143
}$ & $2^{30.132
}$ & $2^{29.769
}$ & $2^{30.138
}$ \\
    cyclic-8       & full     & $2^{31.978
}$ & $2^{31.959
}$ & $2^{31.979
}$ & $2^{30.698
}$ & $2^{30.650
}$ & $2^{30.701
}$ \\
                   & selective& $2^{30.864
}$ & $2^{30.821
}$ & $2^{30.866
}$ & $2^{29.788
}$ & $2^{29.695
}$ & $2^{29.793
}$ \\ \hline
                   & only-top & $2^{23.162
}$ & $2^{22.849
}$ & $2^{23.661
}$ & $2^{22.403
}$ & $2^{21.843
}$ & $2^{23.161
}$ \\
    eco-10         & full     & $2^{26.466
}$ & $2^{26.457
}$ & $2^{26.480
}$ & $2^{23.914
}$ & $2^{23.860
}$ & $2^{23.993
}$ \\
                   & selective& $2^{23.504
}$ & $2^{23.431
}$ & $2^{23.607
}$ & $2^{23.017
}$ & $2^{22.914
}$ & $2^{23.160
}$ \\ \hline
                   & only-top & $2^{26.551
}$ & $2^{26.256
}$ & $2^{26.932
}$ & $2^{25.409
}$ & $2^{24.679
}$ & $2^{26.136
}$ \\
    eco-11         & full     & $2^{30.544
}$ & $2^{30.540
}$ & $2^{30.549
}$ & $2^{27.051
}$ & $2^{27.002
}$ & $2^{27.100
}$ \\
                   & selective& $2^{26.844
}$ & $2^{26.787
}$ & $2^{26.900
}$ & $2^{26.044
}$ & $2^{25.944
}$ & $2^{26.141
}$ \\ \hline
                   & only-top & $2^{12.207
}$ & $2^{12.058
}$ & $2^{13.219
}$ & $2^{11.987
}$ & $2^{11.812
}$ & $2^{13.022
}$ \\
    f-633          & full     & $2^{12.554
}$ & $2^{12.513
}$ & $2^{12.608
}$ & $2^{12.112
}$ & $2^{12.057
}$ & $2^{12.186
}$ \\
                   & selective& $2^{12.187
}$ & $2^{12.134
}$ & $2^{12.257
}$ & $2^{12.064
}$ & $2^{12.006
}$ & $2^{12.140
}$ \\ \hline
                   & only-top & $2^{20.627
}$ & $2^{20.098
}$ & $2^{20.762
}$ & $2^{19.945
}$ & $2^{19.005
}$ & $2^{20.153
}$ \\
    f-744          & full     & $2^{21.094
}$ & $2^{21.084
}$ & $2^{21.128
}$ & $2^{20.284
}$ & $2^{20.267
}$ & $2^{20.344
}$ \\
                   & selective& $2^{20.275
}$ & $2^{20.258
}$ & $2^{20.336
}$ & $2^{19.192
}$ & $2^{19.156
}$ & $2^{19.318
}$ \\ \hline
                   & only-top & $2^{29.633
}$ & $2^{27.907
}$ & $2^{30.009
}$ & $2^{29.548
}$ & $2^{27.635
}$ & $2^{29.926
}$ \\
    katsura-11     & full     & $2^{32.170
}$ & $2^{32.162
}$ & $2^{32.175
}$ & $2^{31.969
}$ & $2^{31.961
}$ & $2^{31.976
}$ \\
                   & selective& $2^{29.932
}$ & $2^{29.897
}$ & $2^{29.958
}$ & $2^{29.907
}$ & $2^{29.871
}$ & $2^{29.934
}$ \\ \hline
                   & only-top & $2^{20.402
}$ & $2^{20.401
}$ & $2^{23.998
}$ & $2^{19.802
}$ & $2^{19.801
}$ & $2^{19.801
}$ \\
    noon-8         & full     & $2^{22.128
}$ & $2^{22.128
}$ & $2^{22.279
}$ & $2^{22.231
}$ & $2^{22.231
}$ & $2^{22.373
}$ \\
                   & selective& $2^{22.169
}$ & $2^{22.169
}$ & $2^{22.316
}$ & $2^{22.059
}$ & $2^{22.059
}$ & $2^{22.218
}$ \\ \hline
                   & only-top & $2^{23.198
}$ & $2^{23.198
}$ & $2^{27.216
}$ & $2^{22.394
}$ & $2^{22.393
}$ & $2^{27.110
}$ \\
    noon-9         & full     & $2^{25.029
}$ & $2^{25.029
}$ & $2^{25.220
}$ & $2^{25.140
}$ & $2^{25.140
}$ & $2^{25.317
}$ \\
                   & selective& $2^{25.076
}$ & $2^{25.076
}$ & $2^{25.261
}$ & $2^{24.969
}$ & $2^{24.969
}$ & $2^{25.167
}$ \\ \hline
                   & only-top & $2^{27.035
}$ & $2^{26.085
}$ & $2^{27.103
}$ & $2^{27.036
}$ & $2^{26.087
}$ & $2^{27.104
}$ \\
    Random(10,2,2) & full     & $2^{25.936
}$ & $2^{25.872
}$ & $2^{25.984
}$ & $2^{25.937
}$ & $2^{25.874
}$ & $2^{25.985
}$ \\
                   & selective& $2^{26.038
}$ & $2^{25.979
}$ & $2^{26.083
}$ & $2^{26.039
}$ & $2^{25.980
}$ & $2^{26.084
}$ \\ \hline
                   & only-top & $2^{29.967
}$ & $2^{29.024
}$ & $2^{30.022
}$ & $2^{29.968
}$ & $2^{29.025
}$ & $2^{30.023
}$ \\
    Random(11,2,2) & full     & $2^{28.831
}$ & $2^{28.770
}$ & $2^{28.867
}$ & $2^{28.831
}$ & $2^{28.770
}$ & $2^{28.867
}$ \\
                   & selective& $2^{28.937
}$ & $2^{28.880
}$ & $2^{28.970
}$ & $2^{28.938
}$ & $2^{28.882
}$ & $2^{28.972
}$ \\ \hline
  \end{tabular}
  \end{center}
\end{table}

\begin{table}[htb]
\begin{center}
  \caption{The numbers of generated S-pairs which satisfy {\fbox{{\bf SF}}} and does not satisfy {\fbox{{\bf SF}}} (homogeneous)} 
  \begin{tabular}{|l||r|r||r|r|} \hline    
    \multicolumn{1}{|l|}{}
      %& \multicolumn{1}{|r|}{}
      & \multicolumn{2}{c|}{ADD} 
      & \multicolumn{2}{c|}{RAT} \\ \cline{2-5}
    \multicolumn{1}{|l|}{benchmark}
      & \multicolumn{1}{c|}{\; {\bf SF} \;} 
      & \multicolumn{1}{c|}{not {\bf SF}}
      & \multicolumn{1}{c|}{\; {\bf SF} \;} 
      & \multicolumn{1}{c|}{not {\bf SF}}\\ \hline \hline
    cyclic-7   & 477
  & 465
  & 477
  & 265
  \\ \hline
    cyclic-8   & 1515
 & 4011
 & 1515
 & 2342
 \\ \hline
    eco-10     & 417
  & 507
  & 417
  & 135
  \\ \hline
    eco-11     & 844
 & 1517
  & 844
 & 303
  \\ \hline
    f-633      & 46
   & 7
   & 46
   & 2
   \\ \hline
    f-744      & 380
  & 354
  & 380
  & 158
  \\ \hline
  katsura-11   & 884
 & 1293
  & 884
 & 1245
  \\ \hline
    noon-8     & 1336
 & 40
 & 1336
 & 40
 \\ \hline
    noon-9     & 3680
 & 54
 & 3680
 & 54
 \\ \hline
HRandom(10,2,2)& 778
  & 144
  & 778
  & 144
  \\ \hline
HRandom(11,2,2)& 1479
 & 342
 & 1479
 & 342
 \\ \hline
  \end{tabular}
\end{center}
\end{table}

\begin{table}[htb]
\begin{center}
  \caption{The numbers of generated S-pairs which satisfy {\fbox{{\bf SF}}} and does not satisfy {\fbox{{\bf SF}}} (inhomogeneous)} 
  \begin{tabular}{|l||r|r||r|r|} \hline    
    \multicolumn{1}{|l|}{}
      %& \multicolumn{1}{|r|}{}
      & \multicolumn{2}{c|}{ADD} 
      & \multicolumn{2}{c|}{RAT} \\ \cline{2-5}
    \multicolumn{1}{|l|}{benchmark}
      & \multicolumn{1}{c|}{\; {\bf SF} \;} 
      & \multicolumn{1}{c|}{not {\bf SF}}
      & \multicolumn{1}{c|}{\; {\bf SF} \;} 
      & \multicolumn{1}{c|}{not {\bf SF}}\\ \hline \hline
    cyclic-7   & 475
  & 467
 & 475
 & 267
 \\ \hline
    cyclic-8   & 1504
 & 4022
 & 1504
 & 2353
 \\ \hline
    eco-10     & 315
  & 629
 & 315
 & 130
 \\ \hline
    eco-11     & 634
 & 1849
 & 634
 & 283
 \\ \hline
    f-633      & 46
   & 7
 & 46
 & 2
 \\ \hline
    f-744      & 333
  & 229
 & 333
 & 160
 \\ \hline
  katsura-11   & 884
 & 1293
 & 884
 & 1245
 \\ \hline
    noon-8     & 1336
 & 40
 & 1336
 & 40
 \\ \hline
    noon-9     & 3680
 & 54
 & 3680
 & 54
 \\ \hline
 Random(10,2,2)& 778
  & 144
 & 778
 & 144
 \\ \hline
 Random(11,2,2)& 1479
 & 342
 & 1479
 & 342
 \\ \hline
  \end{tabular}
\end{center}
\end{table}

\end{document}